\documentclass[%
 reprint,
superscriptaddress,
showpacs,
 amsmath,amssymb,
 aps,
]{revtex4-1}

\usepackage{graphicx}
\usepackage{color}
\usepackage{verbatim}
\usepackage{subfigure}
\DeclareMathOperator{\sech}{sech}

\definecolor{Blue}{rgb}{0,0,1}
\definecolor{Red}{rgb}{1,0,0}
\definecolor{Black}{rgb}{0,0,0}
\newcommand{\DB}[1]{{\color{Black} #1}}

\begin{document}

\preprint{APS/123-QED}

\title{Zero supermode-based multipartite entanglement in $\chi^{(2)}$ nonlinear waveguides arrays} 

\author{David Barral} 
\email{Corresponding author: david.barral@c2n.upsaclay.fr}
\author{Kamel Bencheikh}
\author{Nadia Belabas}
\author{Juan Ariel Levenson}
\affiliation{Centre de Nanosciences et de Nanotechnologies C2N, CNRS, Universit\'e Paris-Saclay, 10 Boulevard Thomas Gobert, 91120 Palaiseau, France}

\begin{abstract} We show that arrays of $\chi^{(2)}$ nonlinear waveguides in the second harmonic generation regime are a promising source of continuous-variable entanglement. We indeed demonstrate analytically that optical arrays with odd number of waveguides injected  with the zero-eigenvalue fundamental supermode entangle this fundamental supermode with a collective harmonic field. Moreover the fundamental individual modes are multipartite entangled and their entanglement grows with propagation length. The device is scalable, robust to losses, does not rely on specific values of nonlinearity and coupling and is easily realized with current technology. It thus stands as an unprecedented candidate for generation of multipartite continuous-variable entanglement for optical quantum information processing.
\end{abstract}

\date{April 14, 2019}
\maketitle 

Einstein, Podolsky and Rosen's (EPR) celebrated gedanken experiment paid attention on the nonlocality of quantum mechanics by considering the case of two spatially separated quantum particles that have both maximally correlated momenta and maximally anticorrelated positions \cite{Reid2009}. Besides the philosophical implications of that work, it gave rise to the concept of quantum entanglement, which underpins current quantum technology \cite{Acin2018}. Remarkably, that paradigmatic example dealt with continuous variables (CV), i.e. variables that can take a continuous spectrum of eigenvalues \cite{Furusawa2015}. Nowadays, CV-based quantum information can be encoded in the fluctuations of the optical-field quadratures and entanglement has extended {from bipartite} to multipartite systems. Recent table-top experiments have demonstrated multipartite CV entanglement in the spatial, frequency and temporal domains \cite{Armstrong2015, Roslund2013, Yoshikawa2016}. However, scalability, stability and transfer to real technologies are  milestones far from feasible with bulk-optics systems. Integrated optics is a leading substrate technology for real-world light-based quantum information technologies: miniaturization, subwavelength stability, and generation, manipulation and detection of entanglement {have recently become} available on chip in the discrete variable regime where individual photons are usually considered \cite{Tanzilli2012}. In the CV regime, bipartite entanglement has very recently been demonstrated on chip in a non scalable scheme  \cite{Lenzini2018}. In this paper we present a simple and practical protocol for the generation of spatial multipartite CV entangled states of light on chip. Bipartite and tripartite CV entanglement has been predicted in arrays of nonlinear waveguides in the spontaneous and stimulated parametric downconversion regime \cite{Rai2012}. However, in that configuration tripartite entanglement is only produced for critical values of the involved parameters. We consider here the case of a $\chi^{(2)}$ nonlinear waveguide array with an odd number of waveguides in the second harmonic generation (SHG) regime. We demonstrate that the zero-eigenvalue fundamental supermode of the equivalent linear array is squeezed along propagation leading to multipartite entanglement between the individual modes. We have found analytical solutions to this system and our method is scalable for any odd number of waveguides. The scheme relies on coupling and nonlinearity within the array but not on the specific values of the parameters, which makes it all the more robust and attractive. The necessary technology to implement this protocol is currently available: discrete quadratic solitons, competing nonlinearities and nonclassical biphoton states have been demonstrated in periodically poled lithium niobate (PPLN) arrays in the last few years \cite{Iwanow2004, Setzpfandt2009, Solntsev2014}. We point out that this work presents the first rigorous demonstration of i) an analytical dynamical solution for SHG in waveguide arrays and ii) scalable integrated CV multipartite entanglement.

 \begin{figure}[t]
  \centering
    {\includegraphics[width=0.46\textwidth]{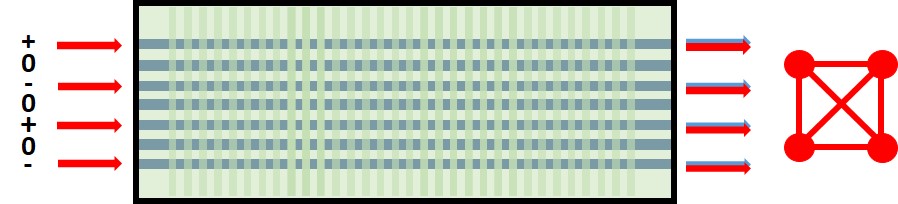}}
\vspace {0cm}\,
\hspace{0cm}\caption{\label{F4}\small{Sketch of the zero-eigenvalue fundamental supermode propagation in a quadratic nonlinear waveguide array with an odd number of waveguides, seven in this example. The numbers on the left indicate the relative magnitude of the input field amplitudes. In red the fundamental modes (FF). In blue the harmonic generated modes (SHF). The nodes are the individual fundamental modes, the vertices stand for entanglement. The maximally connected square indicates genuine four-partite entanglement.}}
\end{figure}

We consider an array of $\chi^{(2)}$ nonlinear waveguides, sketched in Figure 1, made of an odd number $N$ of identical $\chi^{(2)}$ waveguides. \DB{These nonlinear waveguides can be created in noncentrosymmetric materials such as for instance lithium niobate or potassium tytanyl phosphate through proton or ion exchange, diffusion and etching. The strong confinement of the propagating fields produces a high nonlinear interaction \cite{Alibart2016}.} In each waveguide, an input fundamental field (FF) at frequency $\omega_{f}$ is up-converted into a second-harmonic field (SHF) at frequency $\omega_{h}$. We assume that the phase matching condition  $\Delta \kappa \equiv \kappa(\omega_{h})-2 \kappa(\omega_{f})=0$, with $\kappa(\omega_{f,h})$ the propagation constant at frequency $\omega_{f,h}$, is fulfilled only in the coupling zone. The energy of the propagating fundamental modes is exchanged between the coupled waveguides through evanescent waves, whereas the interplay of the generated second harmonic waves is negligible for the considered propagation lengths due to their high confinement into the guiding region. We consider an homogeneous nonlinear array \DB{and continuous-wave propagating fields}. The physical processes involved are described by the following system of equations \cite{Linares2008}
\begin{align} \nonumber
&\frac{d \hat{A}_{f,j}}{d z}=\, i C (\hat{A}_{f,j-1}+\hat{A}_{f,j+1}) +2 i g \hat{A}_{h,j} \hat{A}_{f,j}^{\dag}, \\
\label{QE1}
&\frac{d \hat{A}_{h,j}}{d z}=\, i g \hat{A}_{f,j}^{2}, 
\end{align}
where $\hat{A}_{f,0}=0$ and $\hat{A}_{f,N+1}=0$, and $j=1,\dots, N$ is the individual mode index. \DB{$\hat{A}_{f (h),j}\equiv \hat{A}_{j}(z, \omega_{f (h)})$ are monochromatic slowly-varying amplitude annihilation operators of fundamental (f) and second harmonic (h) photons corresponding to the j-th waveguide 
where $[\hat{A}_{j}(z,\omega),\hat{A}^{\dag}_{j'}(z',\omega')]=\delta(z-z') \delta(\omega-\omega') \delta_{j,j'}$.} $g$ is the nonlinear parameter proportional to $\chi^{(2)}$ and the overlap of the FF and SHF in each waveguide, $C$ the linear coupling constant between neighboring waveguides assumed constant, and $z$ is the coordinate corresponding to the direction of propagation. $C$ and $g$ are taken as real without loss of generality.

To solve the system of Equations (\ref{QE1}), we apply the linearization method by means of quantum-fluctuation operators $\hat{a}_{f(h),j}=\hat{A}_{f(h),j}-\alpha_{f(h),j}$ with $\alpha_{f(h),j}$ the mean values corresponding to the input operators $\hat{A}_{f(h),j}$ \cite{Ou1994}. These new operators exhibit zero mean values and the same variances as the input ones. It is natural to use here the fundamental supermode (normal) basis which diagonalizes the linear part of Equations $(\ref{QE1})$. This change of basis corresponds to the transformation $\hat{\boldsymbol{a}}_{f}=\mathbf{M}\,\hat{\mathbf{b}}_{f}$, where $\hat{\boldsymbol{a}}_{f}=(\hat{a}_{f,1}, \dots, \hat{a}_{f,N})^{T}$ and $\hat{\mathbf{b}}_{f}=(\hat{b}_{f,1}, \dots, \hat{b}_{f,N})^{T}$ are annihilation operator column vectors in the individual and supermode basis, respectively, and $\mathbf{M}$ is the transformation matrix with elements given by \cite{Kapon1984}
\begin{equation}\label{Transf}
{M}_{j,k}=\frac{\sin(\frac{j k \pi}{2l})}{\sqrt{l}},
\end{equation} 
with $l=(N+1)/2$ and $k=1,\dots, N$ is the supermode index. This is a real orthogonal matrix $\mathbf{M}=\mathbf{M}^{-1}$ that operates on quantum operators but also on the classical amplitudes. The propagation constants of the slowly varying supermodes are the eigenvalues of $\mathbf{M}$: $\lambda_{k}=2 C \cos(k \pi/2l)$. Notably when $k=l$, $\lambda_{l}=0$. Thus, the $l$-th supermode presents a zero eigenvalue and does not undergo discrete diffraction \cite{Iwanow2005}. This supermode only appears in arrays with an odd number of waveguides \cite{Efremidis2005}. 

Under the linearization approximation, the propagation of the classical fields $\alpha_{f(h),j}$ is firstly solved to obtain the evolution of the quantum fluctuations. Here, we use the fundamental supermode basis to find the solutions of the classical propagation equations for the amplitudes $\alpha_{h,j}$ and $\beta_{f,k}$, where $\beta_{f,k}$ is the classical mean value of the $k$-th supermode. In the SHG regime, the harmonics initial conditions are $\alpha_{h,j}(0)=0$. We find that under the initial condition $\beta_{f,k}(0)=\delta_{k,l}$, i.e. pumping with the fundamental $l$-th supermode of zero eigenvalue, the propagation Equations (\ref{QE1}) are equivalent to those related to a single waveguide given by (see supplemental mat. \cite{Note0})
\begin{align} \nonumber
&\frac{d {\beta}_{f,l}}{d z}= 2ig  \,\alpha_{h,l}\,{\beta}_{f,l}^{*} ,\\ \label{QS1}
&\frac{d \alpha_{h,2j-1}}{d z}=\frac{d \alpha_{h,l}}{d z}= \frac{ig}{l} \beta_{f,l}^{2},
\end{align}
where we have used the fact that harmonic fields are only generated in the odd waveguides and all have the same evolution. Notably, the supermode $\beta_{f,l}$ remains phase-matched along propagation. Indeed, since the amplitudes of the non-zero input supermode components are equal, the amplitudes of the harmonic fields generated in the odd waveguides are also equal, and fundamental and harmonic fields remain phase-matched. The other supermodes $\beta_{f,m}$ where $m \neq l$ produce nonlinear-based detrimental phases which lead to supermodes coupling and make the system of Equations (\ref{QE1}) non integrable in general \cite{Bang1997}. In contrast, $\beta_{f,l}$ does not produce any nonlinear-based cascade phase \cite{Barral2017, Barral2018}. To our knowledge, this is the first time analytical dynamical solutions are presented for a system composed of an arbitrary odd number $N$ of coupled nonlinear waveguides in the SHG regime. This is our first result. Note that the appropriate initial conditions can be realized by means of off-the-shelf elements like fiber attenuators, phase shifters and V-groove fiber arrays.

\begin{figure}[t]
  \centering
    {\includegraphics[width=0.48\textwidth]{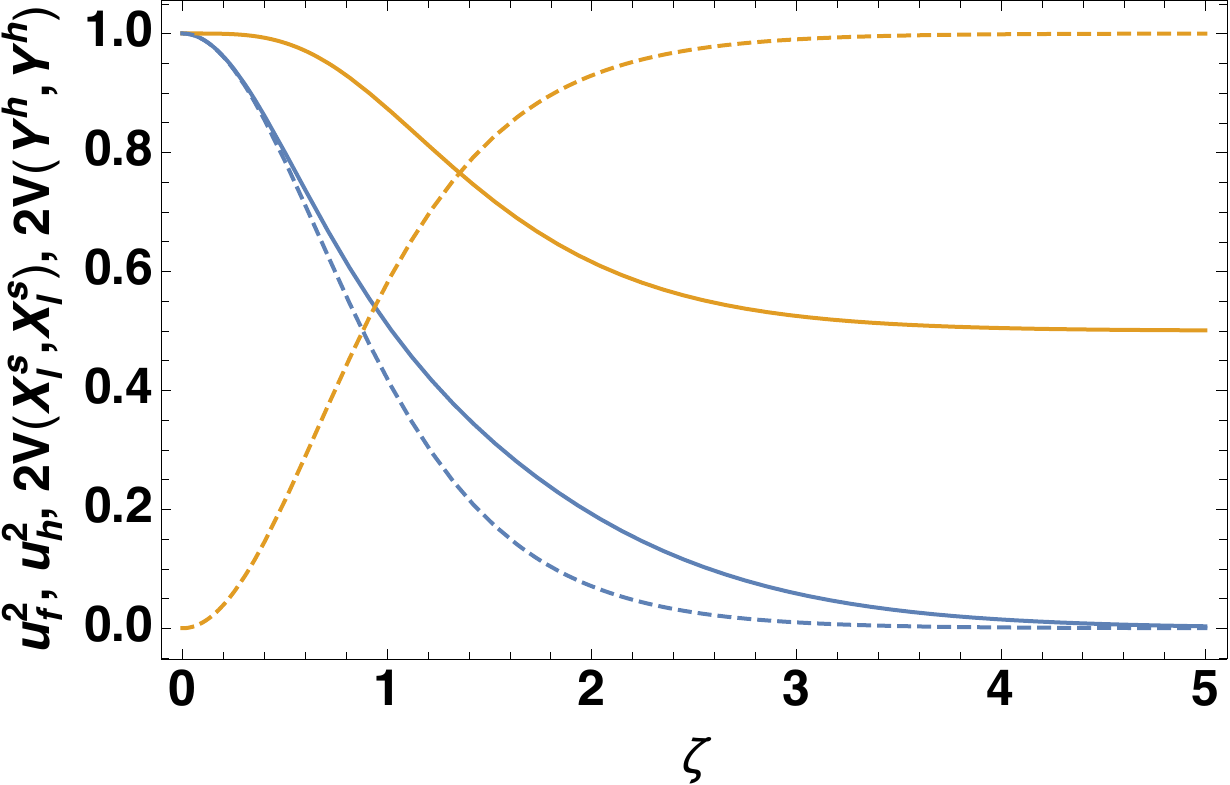}}
\vspace {0cm}\,
\hspace{0cm}\caption{\label{F2}\small{Classical fields power propagation and superquadratures squeezing after injection of the $l$-th fundamental supermode. Dimensionless fundamental supermode (blue, dashed) and second harmonic fields (yellow, dashed) powers. Fundamental amplitude superquadrature squeezing (blue, solid) and harmonic phase superquadrature squeezing (yellow, solid). The squeezing plots are normalized to the shot noise. $\zeta$ is the normalized propagation coordinate.}}
\end{figure}

In order to solve Equations (\ref{QS1}), we use dimensionless amplitudes and phases related to the classical fields through $\beta_{f,l}=\sqrt{P} \,u_{f} \exp{(i\, \theta_{f})}$, $\alpha_{h,l}=\sqrt{P/2l} \,u_{h} \exp{(i\,\theta_{h})}$, with $P\equiv\vert \beta_{f,l}(0)\vert^{2}=\vert \beta_{f,l}\vert^{2}+2l \vert \alpha_{h,l} \vert^{2}$ the total energy in the device. We introduce the energy per waveguide $P_{l}$ as $P_{l}\equiv P/l$, and the normalized propagation coordinate $\zeta=\sqrt{2 P_{l}} \,g z$, which is defined only in the coupling region where phase matching is guaranteed. 
Applying this change of variables into Equations (\ref{QS1}), we obtain for the modes propagating in the nonlinear array
\begin{align}\nonumber
\frac{d {u}_{f}}{d \zeta}=&  - u_{f} u_{h} \sin(\Delta\theta),  \, &\frac{d {\theta}_{f}}{d \zeta}=\,  u_{h} \cos(\Delta\theta),
 \\ \label{us1}
\frac{d {u}_{h}}{d \zeta}=& \,u_{f}^2 \sin(\Delta\theta), & \frac{d {\theta}_{h}}{d \zeta}=\, \frac{u_{f}^2}{u_{h}} \cos(\Delta\theta),
\end{align}
with $\Delta\theta \equiv \theta_{h}-2\theta_{f}$. This system has the well known solutions given by \cite{Armstrong1962}
\begin{align}\label{Usol}
u_{f}(\zeta)=&\sech(\zeta),  &\theta_{f}(\zeta)=0, \nonumber \\  
u_{h}(\zeta)=&\tanh(\zeta),  &\theta_{h}(\zeta)=\pi/2, 
\end{align}
where we have chosen $\theta_{f}(0)=0$ as global input phase of the supermode. Figure \ref{F2} shows dimensionless classical powers for the fundamental $l$-th supermode (blue, dashed) and one harmonic mode (yellow, dashed). The energy efficiently transfers from the fundamental supermode to the harmonic fields and full SHG conversion is obtained for long $\zeta$. Energy conservation $u_{f}^{2}+u_{h}^{2}=1$ is satisfied all along the propagation.

The solutions of the classical system of equations are then fed into first-order equations in the quantum fluctuations keeping only the linear terms. We are mainly interested in the CV quantum noise features of the fundamental modes. Thus we study the evolution of the amplitude (and phase) superquadratures related to fundamental $l$-th supermode $\hat{X}_{l}^{s}$($\hat{Y}_{l}^{s})=\sum_{2j-1=1}^{l} M_{l,2j-1}\hat{X}_{2j-1}^{f}$($\hat{Y}_{2j-1}^{f})$, where $[\hat{X}_{l}^{s},\hat{Y}_{l}^{s}]=i$. The harmonic fields do not have supermodes. However, we use a linear combination of individual harmonic quadratures, or 
effective harmonic superquadratures,  $\hat{X}_{}^{h}$($\hat{Y}_{}^{h})=(1/\sqrt{l})\sum_{2j-1=1}^{l} \hat{X}_{2j-1}^{h}$($\hat{Y}_{2j-1}^{h})$ \cite{Note1}, where $[\hat{X}^{h},\hat{Y}^{h}]=i$, to capture the propagation of the quantum field quadratures in a  simple set of equations
\begin{align} \label{Quad1}
\frac{d \hat{X}^{s}_{l}}{d \zeta}= &-\tanh(\zeta)\hat{X}^{s}_{l}-\sqrt{2} \sech(\zeta) \hat{Y}^{h}, \nonumber \\ 
\frac{d \hat{Y}^{s}_{l}}{d \zeta}=  & \tanh(\zeta)\hat{Y}^{s}_{l}+\sqrt{2} \sech(\zeta) \hat{X}^{h},   \nonumber \\ 
\frac{d \hat{X}^{h}}{d \zeta}=    &  -\sqrt{2} \sech(\zeta) \hat{Y}^{s}_{l},    \nonumber \\
\frac{d \hat{Y}^{h}}{d \zeta}=  &\,   \sqrt{2} \sech(\zeta) \hat{X}^{s}_{l}.  \end{align}
The solution of this system of equations is given by $\hat{\xi}^{s}(\zeta)=\mathbf{U}^{s}(\zeta)\, \hat{\xi}^{s}(0)$, where $\hat{\xi}^{s}=(\hat{X}^{s}_{l},\hat{Y}^{s}_{l},\hat{X}^{h},\hat{Y}^{h})^T$ stands for the superquadratures and the evolution operator is given by
\begin{equation} \label{US}
\mathbf{U}^{s}(\zeta)=\begin{pmatrix}
s_{s}^{x} & 0 & 0 & s_{h}^{y} \\
0 & s_{s}^{y} & s_{h}^{x} & 0 \\
0 & h_{s}^{y} & h_{h}^{x} & 0 \\
h_{s}^{x} & 0 & 0 & h_{h}^{y} 
\end{pmatrix},
\vspace{0.05cm}
\end{equation}
with $s^{x}_{s}=\sech(\zeta)(1-\zeta\tanh(\zeta))$, $s_{s}^{y}=\sech(\zeta)$, $s_{h}^{x}=(\sinh(\zeta)+\zeta\sech(\zeta))/\sqrt{2}$, $s_{h}^{y}=-\sqrt{2} \tanh(\zeta) \sech(\zeta)$, $h_{s}^{x}=(\tanh(\zeta)+\zeta\sech^{2}(\zeta))/\sqrt{2}$, $h_{s}^{y}=-\sqrt{2} \tanh(\zeta)$, $h_{h}^{x}=1-\zeta\tanh(\zeta)$ and $h_{h}^{y}=\sech^{2}(\zeta)$. The amplitude (phase) fundamental superquadrature couples only to the phase (amplitude) harmonic superquadrature. We obtain in our multimode system a result similar to what was obtained in ref. \cite{Ou1994} for one fundamental mode but we use collective quadratures \cite{Note2}.

We now consider the quantum properties of this system. We deal with vacuum, coherent or squeezed states. Thus, experimentally, the most interesting observables for these Gaussian states in terms of their CV features are the second-order moments of the quadrature operators, which are the elements of the covariance matrix $\tilde{\mathbf{V}}$: $\tilde{V}({\xi}_{i}, {\xi}_{j})=\frac{1}{2}(\langle\Delta\hat{\xi}_{i} \Delta\hat{\xi}_{j}\rangle + \langle\Delta\hat{\xi}_{j} \Delta\hat{\xi}_{i}\rangle)$, with $\Delta \hat{\xi}\equiv\hat{\xi}-\langle\hat{\xi} \rangle$ \cite{Adesso2014}. $\tilde{\mathbf{V}}$ is a real symmetric matrix that contains all the useful information about quantum fields correlations and it can be efficiently measured by means of homodyne detection \cite{Dauria2009} or quasiresonant analysis cavities in the case of bright beams \cite{Coelho2009}. The covariance matrix at any normalized propagation plane $\zeta$ for Gaussian input fields is given in general by 
\begin{equation}\label{Cov}
\tilde{\mathbf{V}}(\zeta)=(1/2) \tilde{\mathbf{U}}(\zeta) \,\tilde{\mathbf{U}}^{T}(\zeta),
\end{equation}
with 1/2 the shot noise in our convention and $\tilde{\mathbf{U}}$ the evolution operator in a given basis, either individual or supermodes. 

Applying Equation (\ref{US}) into Equation (\ref{Cov}) we obtain analytical solutions $\mathbf{V}^{s}$ in the superquadratures basis ${\xi}^{s}$ (see supplemental material). Figure \ref{F2} displays the evolution of normalized fundamental amplitude superquadrature squeezing (blue, solid) and harmonic phase superquadrature squeezing (yellow, solid). As $\zeta$ increases, the supermode amplitude squeezing is unlimited whereas the maximum harmonic phase squeezing is limited to $50\%$ \cite{Ou1994}. Remarkably, this solution leads to our second result: the fundamental supermode noise is collectively squeezed along propagation in the waveguide array for any odd number of waveguides. It should be noted here that the phase quadrature fluctuations of the fundamental supermode diverge exponentially limiting the range of validity of the linearization approximation. We have checked that for typical total powers $P_{l}$ of hundreds of $mW$ at telecom wavelengths, the linearization is safe for $\zeta \leq 6$ \cite{Olsen2000}.
 \begin{figure}[t]
  \centering
    {\includegraphics[width=0.48\textwidth]{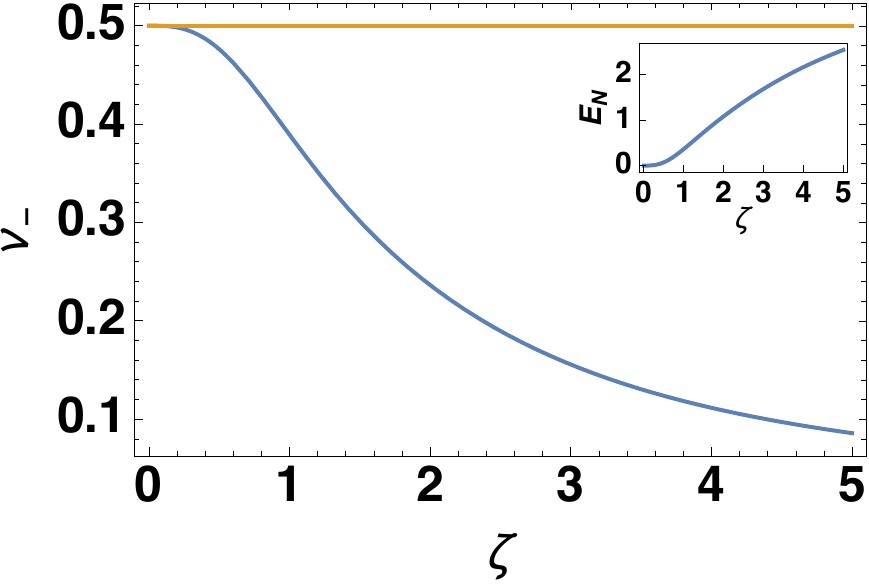}}
\vspace {0cm}\,
\hspace{0cm}\caption{\label{F3}\small{Two-color bipartite entanglement. Fundamental-Harmonic superquadratures entanglement (blue, solid). $\nu_{-}<1/2$  indicates entanglement. \DB{Inset: $E_{N}>0$ indicates entanglement.} $\zeta$ is the normalized propagation coordinate.}}
\end{figure}


Once $\tilde{\mathbf{V}}$ is known, the amount of CV entanglement in bipartite splittings of the system is easily quantified through the Peres-Horodecki-Simon criterion, which establishes that a quantum state is entangled if the partially transposed (PT) density matrix is non-positive. In terms of continuous variables, the entanglement criterion for single-mode bipartite splittings is $\nu_{-}<1/2$, with $\nu_{-}$ the minimum symplectic eigenvalue of the PT-covariance matrix with respect to a subsystem $j$, $\tilde{\mathbf{V}}^{T_{j}}$ \cite{Simon2000}. The closer the value of $\nu_{-}$ to zero, the higher the entanglement between two optical modes (collective or individual). \DB{A well-known entanglement quantifier is the logarithmic negativity $E_{N}$, which is easily computed from the minimum symplectic eigenvalue as $E_{N}=\max[0,-\log_{2} 2\nu_{-}]$ \cite{Vidal2002}. $E_{N}>0$ indicates entanglement. Moreover, $E_{N}$ is a bound of the entanglement of formation $E_{F}$, an entanglement witness with appealing properties from an information-theoretic point of view \cite{Tserkis2017}.} Figure \ref{F3} displays the evolution of two-color entanglement between the fundamental supermode and the collective harmonic field \cite{Note1}. The figure shows that entanglement exists at every $\zeta$ and increases with distance and power. The entanglement is directly related to the partial purities of the subsystems \cite{Adesso2004}. For the fundamental supermode, the partial purity $\mu_{f}$ is $0.97$ at $\zeta=1$ and decreases to $\approx 0.77$ as $\zeta \rightarrow 2$ (see supplemental material), such that it keeps a high value for typical PPLN lengths or power \cite{Kaiser2016}. Notably, no array parameter --coupling or length-- has to be finely set in order to obtain evergrowing entanglement. However, the measurement of two-color entanglement is experimentally demanding \cite{Coelho2009}. The measurement of individual FF quadratures only is best suited here, since the same laser can be used in both generation and detection stages simplifying setups and avoiding problems of mode matching.

We now derive and compute what kind of CV quantum correlations are created intra-supermode, i.e. between the fundamental modes in the individual basis within the $l$-th supermode that steadily grows, squeezes and entangles with the generated harmonics. To that end we trace out the harmonic-mode subsystem and apply the following transformation to the superquadrature covariance matrix
\begin{equation}
\mathbf{V}(\zeta)=\mathbf{M}^{sy} (\frac{1}{2} \mathbf{1}_{f}^{k\neq l})  \mathbf{V}^{s}_{f}(\zeta) (\mathbf{M}^{sy})^{T},
\end{equation}
where $\mathbf{M}^{sy}$ is the symplectic counterpart of the supermode transformation matrix of Equation (\ref{Transf}) \cite{Dutta1995}, $\frac{1}{2} \mathbf{1}_{f}^{k\neq l}$ stands for the shot noise in the supermodes $k\neq l$, and $\mathbf{V}^{s}_{f}$ is the covariance matrix corresponding to the fundamental supermode subsystem. This covariance matrix contains full CV quantum noise information of the system in the basis of the individual modes $\hat{\xi}=(\hat{X}_{1}^{f},\hat{Y}_{1}^{f},\dots, \hat{X}_{N}^{f},\hat{Y}_{N}^{f})^T$ (see supplemental material). Figure \ref{F4} displays the squeezing and entanglement achieved on the individual fundamental modes along propagation in waveguide arrays with $N= 3, 5, 7 $ and $9$ waveguides. More precisely, we obtain a steadily-growing amplitude quadrature squeezing of each of the FF (Figure \ref{F4}\DB{a}) and entanglement between any pair of single-mode propagating FF (Figure \ref{F4}\DB{b}) for typical PPLN lengths or power ($\zeta < 2$). The generated squeezing in the supermode is equally shared between the parties and mixed with the shot noise of the even non-injected channels. This limits the maximum amount of squeezing per party to (see supplemental material)
\begin{equation}\label{SqI}
2V(X^{f},X^{f}) \rightarrow \frac{l-1}{l} =\frac{N-1}{N+1} <1 \quad\forall \,\text{odd N}.
\end{equation}
Thus for a large number of propagating modes the squeezing approaches the shot noise but it is always below it. The entanglement is equal between any pair of fundamental modes propagating in the array. Maximum entanglement is reached for $N=3$ since the squeezing is shared only between the two propagating modes or $l=2$. As the number of parties increases, the available bipartite entanglement is also limited and gets lower. Bipartite entanglement saturates at long distances (or high power) for $N>3$ due to the presence of extra noise in the even channels as (see supplemental material)
\begin{align}\nonumber
\nu_{-}&\rightarrow \frac{1}{2}\sqrt{\frac{l-2}{l}}=\frac{1}{2}\sqrt{\frac{N-3}{N+1}}<\frac{1}{2}, \\ \nonumber
E_{N}&\rightarrow \log_{2}(\sqrt{\frac{l}{l-2}})=\log_{2}(\sqrt{\frac{N+1}{N-3}})>0 \,\,\,\forall \,\text{odd N}\geq3.
\end{align}
Notably, bipartite entanglement is always present independently of the number $N$ of waveguides in the array.

 \begin{figure}[t]
  \centering
    \subfigure{\includegraphics[width=0.48\textwidth]{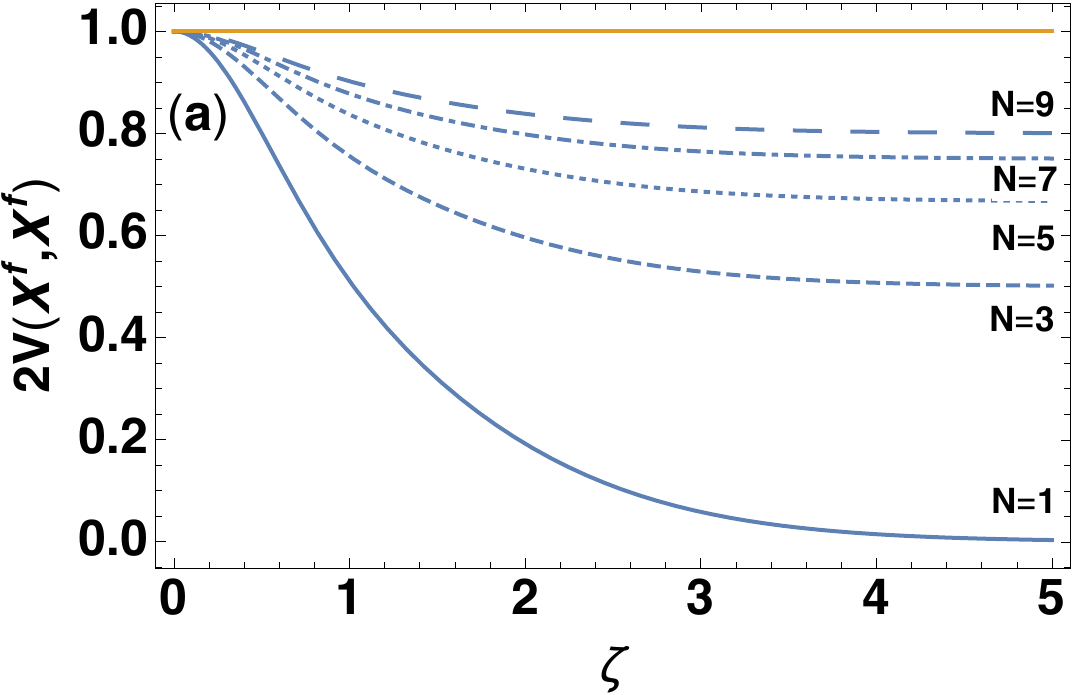}}
    \vspace {-0.2cm}
    \subfigure{\includegraphics[width=0.48\textwidth]{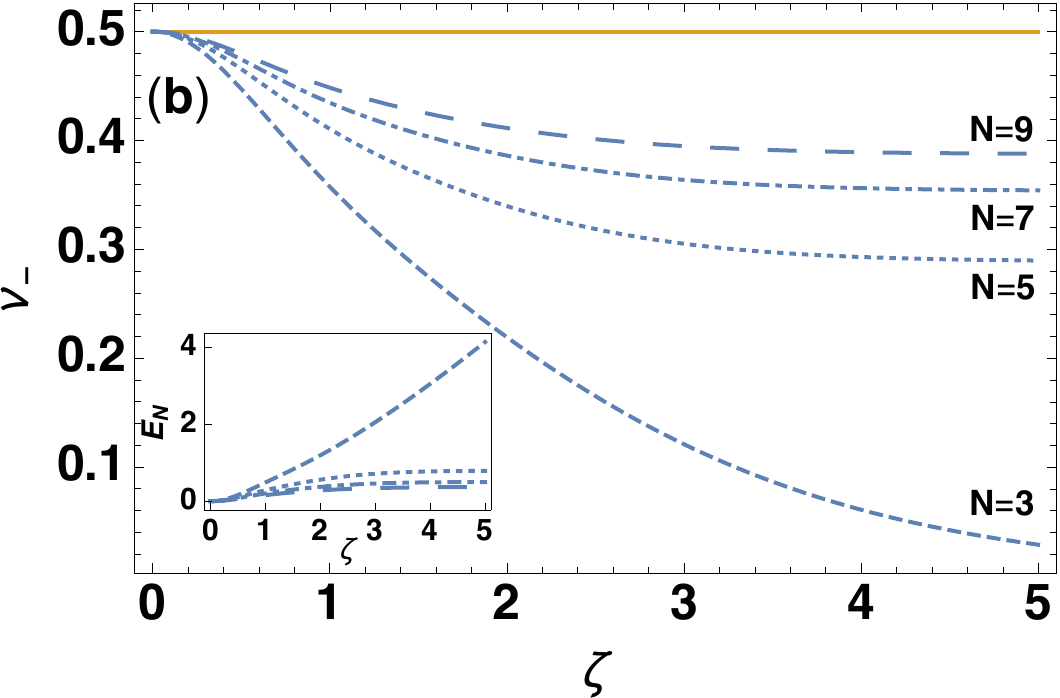}}
\vspace {0cm}\,
\hspace{0cm}\caption{\label{F4}\small{Top \DB{(a)}: fundamental modes amplitude squeezing in waveguide arrays with $N= 1$ (solid), 3 (dashed), 5 (dotted), 7 (dot-dashed) and 9 (large dashed). Normalized shot noise in solid yellow. Bottom \DB{(b)}: bipartite entanglement between fundamental individual modes in waveguide arrays with $N= 3$ (dashed), 5 (dotted), 7 (dot-dashed) and 9 (large dashed). $\nu_{-}<1/2$ indicates entanglement. \DB{Inset: $E_{N}>0$ indicates entanglement.} $\zeta$ is the normalized propagation coordinate.}}
\end{figure}

\begin{figure}[t]
  \centering
    {\includegraphics[width=0.48\textwidth]{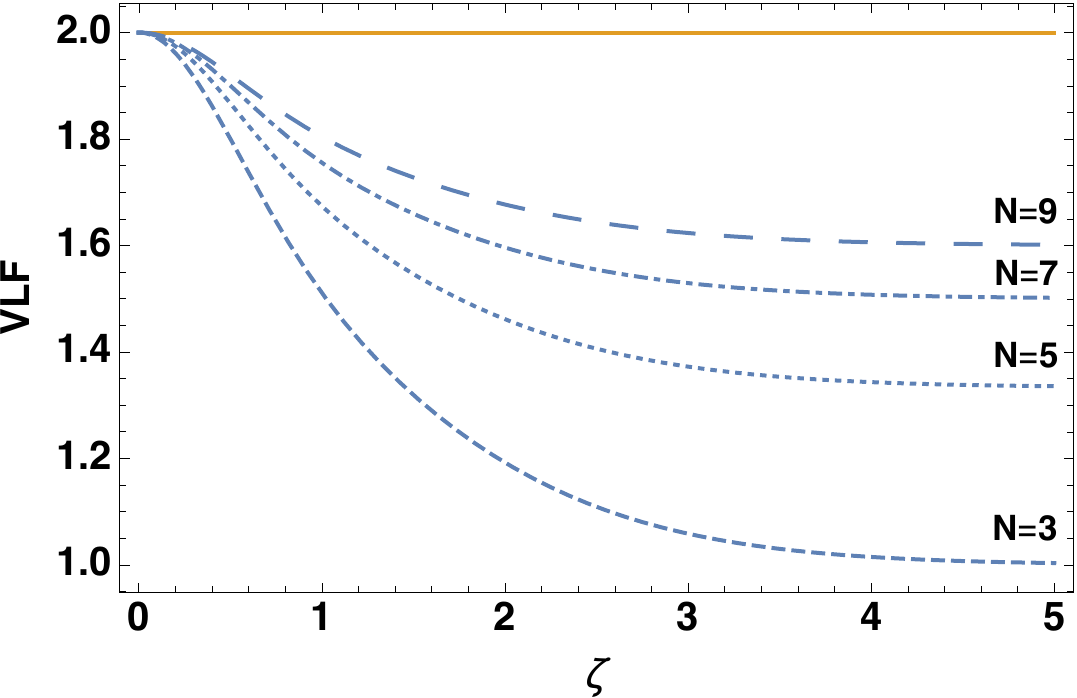}}
\vspace {0cm}\,
\hspace{0cm}\caption{\label{F5}\small{Multipartite entanglement. Optimized van Loock - Furusawa (VLF) inequalities. Simultaneous values under the threshold value VLF=2 (yellow) imply CV tripartite entanglement (two degenerate inequalities, N=5, dotted), quadripartite entanglement (three degenerate inequalities, N=7, dot-dashed) and pentapartite entanglement (four degenerate inequalities, N=9, large dashed). We also show bipartite entanglement for comparison (N=3, dashed). $\zeta$ is the normalized propagation coordinate.}}
\end{figure}

Measuring multipartite full inseparability in CV systems requires the simultaneous fulfillment of a set of conditions which leads to genuine multipartite entanglement when pure states are involved \cite{vanLoock2003,Teh2014}. This criterion, known as van Loock - Furusawa inequalities, can be easily calculated from the elements of the covariance matrix $\mathbf{V}$. Figure \ref{F5} shows two, three and four degenerate inequalities for arrays with, respectively, three ($N=5$), four ($N=7$) and five ($N=9$) propagating modes. We also show bipartite entanglement ($N=3$) for comparison. The optimized violation (VLF$<2$ in our notation) of two, three and four inequalities -- Equations (43) of ref. \cite{vanLoock2003} -- guarantees full inseparability. Since we deal with pure states at the fundamental frequency level (mixed with the collective harmonic mode), the propagating fundamental modes are genuinely multipartite entangled in both cases. The violations are degenerate for each $N$ and saturate at (see supplemental material)
\begin{equation} \nonumber
\text{VLF} = 4V(X^{f},X^{f}) \rightarrow 2\frac{l-1}{l} =2 \frac{N-1}{N+1} <2 \quad\forall \,\text{odd N},
\end{equation}
for large $\zeta$. Thus, the multipartite entanglement is directly related to the squeezing available per mode through Equation (\ref{SqI}). Weakening of quantum correlations as the number of modes increases due to additional vacuum contributions is also found in bulk-optics approaches \cite{Armstrong2015} and, due to practical reasons, the number of vacuum modes is usually much larger than the number of squeezed inputs, preventing scalability. Note that in the case of $N=3$ waveguides the inequality remains above the EPR steering threshold VLF$=1$ \cite{Teh2014} and perfect entanglement VLF$\rightarrow 0$ is never obtained. The VLF criterion thus detects the presence of vacuum in the central waveguide whereas the minimum PT-symplectic eigenvalue $\nu_{-}$ does not. Remarkably, the FF exhibit multipartite entanglement at any $\zeta$ independently of the number of propagating modes and the number of entangled modes scales with $l$, i.e. linearly with the number of waveguides. The multimode entanglement can be improved applying an off-chip distillation protocol by means of non-Gaussian operations \cite{Takahashi2010}. Additionally, this approach minimizes the resources necessary to generate multipartite entanglement in, for instance, telecommunication bands since entanglement is created at the input wavelength \cite{Barral2018}.

Finally, we conclude with a few comments about the feasibility and the range of application of this method. The influence of losses on the CV entanglement can be easily included in our analysis by inserting fictitious beam splitters with a given effective transmittivity \cite{Barral2018}. \DB{Our simulations indicate that propagation losses have a small impact on squeezing and entanglement considering typical values in PPLN waveguides} (see supplemental material) \cite{Lenzini2018}. We have found that the effect of losses is alleviated as the number of waveguides increases. A drop of $\approx 1\%$ in squeezing is obtained at $\zeta=1$ for $N=1$, whereas for $N=9$ this detrimental effect decreases to $\approx 0.3\%$. We also emphasize that for state-of-the-art figures in PPLN waveguides, such as $g=25.10^{-4}$ mm$^{-1}$ mW$^{-1/2}$ and $P_{l}=200$ mW \cite{Lenzini2018, Kaiser2016}, $\zeta=1$ is equivalent to $z=2$ cm. Our method can thus be implemented with current technology. Moreover, a nonlinear efficiency more than an order of magnitude higher is expected in nanophotonic PPLN waveguides \cite{Boes2018, Loncar2018}. 

We have found an analytical dynamical solution for quadratic nonlinear waveguide arrays with odd number of waveguides in the SHG regime. We have demonstrated that this device is a versatile and efficient source of CV entanglement. This configuration is scalable to any dimension, relies on coupling and nonlinearity within the array but not on the specific values of the parameters, it is robust to losses, and the present technology is ready to implement it. Remarkably, our protocol can be extended to other arrays supporting homogeneous zero-eigenvalue supermodes where phase matching is preserved along propagation. This work opens new avenues in the generation of multipartite entangled states through supermodes-supporting devices as multicore fibers and 2D-3D waveguiding structures in optics or Fermi-resonance interface modes in solid state physics \cite{Bang1997}. To conclude, we would like to point out that this is the first full demonstration of CV multipartite entanglement in waveguide arrays. We have found an analytical solution in a usually non-analytical framework. The obtained physical insight on nonlinear waveguides can now be put to good use: to optimize numerically input configurations maximizing squeezing and entanglement by preventing the mixing with the vacuum modes. In a forthcoming paper we will extend this analysis to the spontaneous parametric downconversion regime where careful pump shaping is required to select the desired set of generated fundamental supermodes.

{\it Acknowledgements.} We thank M. Walschaers and N. Treps for useful discussions. This work was supported by the Agence Nationale de la Recherche through the INQCA project \DB{(Grant Agreement No. PN-II-ID-JRP-RO-FR-2014-0013 and ANR-14-CE26-0038)}, the Paris \^Ile-de-France region in the framework of DIM SIRTEQ through the project ENCORE, and the Investissements d'Avenir program (Labex NanoSaclay, reference ANR-10-LABX-0035).

\end{document}